# Zero-depth interfacial nanopore capillaries

*Hadi Arjmandi-Tash[1], Amedeo Bellunato[1], Chenyu Wen[2], René C. Olsthoorn[1], Ralph H. Scheicher[3], Shi-Li Zhang[2] and Grégory F. Schneider[1]\**

[1]Faculty of Science, Leiden Institute of Chemistry, Leiden University, 2333CC Leiden, The Netherlands

[2]Division of Solid-State Electronics, Department of Engineering Sciences, Uppsala University, 75121 Uppsala, Sweden

[3]Division of Materials Theory, Department of Physics and Astronomy, Uppsala University, 75120 Uppsala, Sweden

High-fidelity analysis of translocating biomolecules through nanopores demands shortening the nanocapillary length to a minimal value. Existing nanopores and capillaries, however, inherit a finite length from the parent membranes. Here, we form nanocapillaries of zero depth by dissolving two superimposed and crossing metallic nanorods, thereby opening two overlapping nanofluidic channels molded in a polymeric resin. In an electrolyte, the interface shared by the crossing fluidic channels is mathematically of zero thickness and defines the narrowest constriction in the stream of ions through the nanopore device. This novel architecture provides the possibility to design nanopore fluidic channels, particularly with a robust 3D architecture

Corresponding author: g.f.schneider@chem.leidenuniv.nl

maintaining the ultimate zero thickness geometry independently of the thickness of the fluidic channels. With orders of magnitude reduced biomolecule translocation speed, and lowered electronic and ionic noise compared to nanopores in 2D materials, our findings establish interfacial nanopores as a scalable platform for realizing nanofluidic systems, capable of single-molecule detection.

Conventional nanopores are nano-sized fluidic channels drilled across a solid-state membrane [1–4] or formed by self-assembly of supramolecular compounds [5] and mounted in a flow cell. The flow cell is equally filled with an ionic solution on both sides of the membrane, while a potential difference is applied across the cell serving as the driving force for the ionic transport. Thereby, a flux of ions is established through the nanopore.

Charged molecules can translocate through the nanopore. The instant passage of the molecule momentarily impacts the conductance by locally reducing the aperture size of the channel. The resulting variations of the ionic conductance depends on the local topology of the translocating molecule; particularly, portions of long chain molecules such as polymers, proteins or DNA mark the electronic read-out with specific conductance blockade fingerprints and ultimately allow for reconstructing the sequence of monomers composing the translocating strands [6]. Consequently, thinner pores, i.e. capillaries with shorter channels, are capable of resolving shorter portions of molecules, leading for instance towards high resolution sequencing devices [7]. Thus, the challenge towards high-resolution sequencing has driven the development of ultra-short channel nanopores. Historically, two major classes of nanopores, i.e. biological and solid state nanopores, have been considered. The thickness of these nanopores varies from a few nanometers, as for α-haemolysin biological nanopores [8,9], up to tens of nanometers for solid-state nanopores [10].

A revolutionary breakthrough aiming at reducing the capillary length of nanopores was achieved by the introduction of two-dimensional (2D) materials such as graphene [11–13], hexagonal boron nitride [14] and molybdenum disulfide [15–18]. Indeed, the monoatomic capillary length of 2D nanopores is expected to offer sequencing capabilities [1], but has not been realized yet. Inferior mechanical stability is one of the downside of thin membranes inherently limiting the sustainability of 2D nanopores. Moreover, the complex fabrication process, involving cleanroom facilities and electron beam lithography [19–21], can be demanding to scale up to industrial production. The noise levels in such devices are also orders of magnitude higher than those for long capillary-based nanopores, thus hindering their application for sequencing [22].

To address these issues, we introduce the concept of interfacial nanopores, generated at the crossing of two trenches, as illustrated in Figure 1. Fundamentally, the cross-section of two one-dimensional straight lines is a zero-dimensional entity defined as a point (Figure 1a). The addition of a second dimensionality implies the overlap of two components to become a surface (Figure 1b). Similarly, in a three-dimensional space, the interface shared between two tangent rectangular parallelepipeds is a surface, hence mathematically two-dimensional (Figure 1c). Unlike nanopores commonly fabricated in two-dimensional materials − which notwithstanding still possess a finite thickness − the surface defined by the crossing parallelepipeds is strictly two-dimensional and thus does not exhibit any thickness. A negative mold of this structure therefore yields a nanopore with a capillary of length zero (Figure 1f).

In practice, the rectangular parallelepipeds are fabricated by cutting thin (tens of nanometers in thickness) polymeric slabs containing a gold film (Figure 1d, supplementary information S1) [23]. Positioning two of those slabs on top of each other (Figure 1e) and selectively etching gold yields the interfacial nanopore (Figure 1f). In a typical application, the narrowest constraint in the passage of the buffer solution and ions from one side to the other side of the membrane (two slabs) is of zero thickness. Atomic force micrographs of the fabricated devices bearing the interfacial nanopore in the middle are shown in Figure 1g and 1h, respectively before and after dissolving the gold structures. A glass substrate with a microscale opening at the center was used to provide a mechanical support for the stack of slabs (Figure 1i).

Figure 2a illustrates the I-V characteristic of a nanopore achieved by etching two Au nanorods of 50 nm width and 200 nm height (respectively referred to as *a* and *h* throughout the manuscript, see the inset in Figure 2b), leading to a pore area of 50×50 nm² (see the supplementary information S1 and S2 for the experimental details). The transmembrane potential sweeps from -200 mV to +200 mV and the salt concentration ranges between 1 mM and 1 M. The linear I-V behavior confirms the ionic conduction of a nanopore filled with electrolytic solution and allows to exclude the establishment of any electrochemical reaction within the flow-cell, especially in the proximity of the pore [11–13].

The ionic flow through conventional nanopores experiences a total resistance due to i) the friction with the channel inside the pore region (pore resistance), and ii) the convergence of the electric field lines at the 'mouth' of the nanopore (access resistance). Interestingly, the 2D nature of the interfacial nanopore eliminates the pore resistance term. Still, the access resistance of an interfacial nanopore is composed of two terms: i) the access resistance between the reservoir and the channels in the polymeric slabs, and ii) the resistance inside the channel towards the pore, on both sides. Analytically, the overall access resistance ($R_t$) of an interfacial nanopore is expressed as (see supplementary information S3):

$$R_t = \gamma \frac{F(a,b,h)}{\pi abqc^n(\mu_+ + \mu_-)} \tag{1}$$

where $c$ is the salt concentration in the electrolyte, $q = 1.6 \times 10^{-19}$ C is the elementary charge, $\mu_+$ and $\mu_-$ are the mobility of cations and anions, $a$ and $b$ are the width of the upper and lower channels and $h$ is the equal thickness of the slabs (inset in Figure 2b). $F$ is a function of the geometrical parameters explained in supplementary information S3. The fitting parameters $\gamma$ and $n$ are introduced to take into account the surface conductivity of the pore upon the formation of an electrical double layer, which may impact on the linearity of the I-V curves.

Based on the model in Equation 1, Figure 2b provides a mapping for the expected resistance of the nanopore upon changing the geometrical parameters $a$ and $h$ (here $a = b$). The dependency of the resistance on the trench width is normally stronger than on the slab thickness; particularly for $h > 80\ nm$, the resistance is almost independent of $h$.

We experimentally measured the conductance of interfacial nanopores with different trench widths ranging from 10 nm up to 70 nm (Figure 2c). As expected, increasing $a$ lowers the resistance due to the diffusion of ions leading to increased conductances in widened trenches. In Figure 2c, the continuous lines representing the prediction of the model in Equation 1 match with the experimental results for KCl concentrations above 1mM. As expected, at lower KCl concentrations – and similarly to conventional solid-state nanopores – surface charges on the channel walls yield higher conductances than the one predicted by our model [24,25]. Remarkably and as predicted (Figure 2b), the effect of the slab thickness on the measured ionic resistance is negligible, most particularly for slabs thicker than tens of nanometers for ionic strengths above 10mM (Figure 2d). Again, at lower salt concentrations surface charges add-up to the total conductance of the nanopore architecture.

Figure 3a shows a typical time trace of the ionic current through an interfacial nanopore (a = 70 nm, h = 50 nm) immersed in a 5 mM LiCl buffer solution. Upon addition of 48.5 kbp λ-DNA molecules, a series of drops in the conductance of the nanopore appears, depicting the translocation of DNA molecules through the nanopore. Translocation was verified by a polymerase chain reaction (PCR) experiment (Figure 3a, supplementary information S4).

The duration and blockade current of the translocation events (~400 events) are plotted in Figure 3b. Two highly-populated events with Gaussian distributions are identified in both histograms (green and blue dashed curves) that can be attributed to the translocation of DNA molecules with different foldings. The more populated component exhibits an average translocation duration of ~22ms which corresponds to ~450 ns/bp. Interestingly the measured dwell time is 1.5 to ~100 times longer than the reports for two-dimensional (5 ns/bp [12] - 56 ns/bp [1]), biological (30 ns/bp [26]) and solid-state (40 ns/bp - 300 ns/bp) nanopores [27].

Several observations suggest the presence of a strong interaction between DNA and the walls of the trench, which eventually slows down the translocation of molecules. First, the majority of the translocation events in interfacial nanopores starts sharply but ends smoothly (Figure 3a). This observation can be well explained considering a binding mechanism between DNA and the walls of the trench; in fact, the binding requires time and energy to break, in order to let the DNA exit the nanopore (Figure S5). Second, increasing the salt concentration lowers the dwell time through interfacial nanopores (Figure S4c). This observation is in striking contrast to the reported behavior of DNA in SiNx nanopores [27] in which the strong binding between $Li^+$ to DNA suppresses the translocation speed in high salt concentrations and can be well explained by considering the DNA-nanopore interaction. Third, the widely spread event duration, ranging from less than ~14 ms to over 80 ms (Figure 3b), is another signature of the DNA-nanopore interaction: in the absence of such interaction, DNA molecules are expected to exhibit uniform translocations [28]. Hydrophobic interaction between DNA and the trench walls or cross-over from base-base pi-stacking to base-polymer pi-stacking [29] may govern the DNA-wall interaction.

The ionic resistance of nanopores, generally, is intuitively dominated by that of the most restrictive point (e.g. the interface region for interfacial nanopores) where the electric field is the strongest. Thus, the effective thickness of the nanopore can be defined by referring to the profile of the electric field along the central axis perpendicular to the nanopore area (Figure S6). Specifically, the effective thickness is twice the distance from the nanopore center to the point where the electric field intensity drops to 1/e of its peak value. According to this definition, the effective thickness of interfacial nanopores with varied slab thicknesses is compared with that of conventional nanopores in Figure 3c. We recall the discussion from an earlier section where while ionic resistance in conventional nanopores consists of two components (pore and access resistances), the zero-geometrical thickness (as opposed to the effective thickness) of interfacial nanopores suppresses any contribution of the pore resistance. Indeed, interfacial nanopores show obvious advantages (lower effective nanopore thickness) over conventional nanopores with the channel thicknesses larger than the nanopore size ($h > a$). Our simulations show that the channel length (the thickness of the membrane) governs the effective thickness of conventional nanopores (Figure S6e).

In the other extreme ($h < a$, comparable to the typical geometry of 2D nanopores) the effective thickness on each side of the interfacial nanopores can be estimated as half the width of the trench, $a/2$. This estimation resembles the conventional picture of the access resistance in single circular nanopores as two hemispheres with radius $r = d/2$ (where $d$ is the diameter of the nanopore) at each side of the membrane [30,31]. Here, the interfacial nanopores are clearly advantageous since due to the lack of any pore resistance, its effective thickness always falls below that of the conventional nanopores (sum of the access and pore region, Figure S6f and inset Figure 3c).

Most conventional nanopores operate in membrane thicknesses of ~ 20 nm. As is demonstrated by our simulations (inset and main panel in Figure 3c), a conventional nanopore of h = 4 nm is preferred over the one of h = 100 nm as the former provides an effective thickness of ~8 times smaller (higher resolution); but at the same time such a thin membrane cannot be mechanically as stable as the thick one. Hence a thickness of ~20 nm provides a compromise between the sensitivity and stability. The introduction of interfacial nanopores dramatically shifts this compromise: here the effective thickness of a nanopore with h = 100 nm is just ~1.7 times higher than that of h = 4 nm; hence much thicker nanopores can be chosen without losing the resolution considerably. This is an intriguing property of the interfacial nanopores as the thickness of the membrane and the effective thickness (resolution) are now disentangled. The design of interfacial nanopores is unique as it eliminates the pore thickness; the remaining access resistance term can be minimized by optimizing the geometrical parameters (lowering the area of the pore, Figure 3d). Then the design allows to reach an ultimate resolution which is not reachable with conventional designs, always having a finite pore thickness. We note that the experimental evidences for an ultimate resolution can be achieved only when biomolecule sequencing is performed; this is not the case so far as prominent experimental challenges including high translocation speed of molecules do not allow single base reading [1].

Figure 4a compares the noise power spectral densities (PSD, denoted by $S_I$) of three types of nanopores, including a nanopore in graphene, a nanopore in SiNx and an interfacial nanopore, all of similar ionic conductances and comparable nanopore areas. The parasitic capacitive coupling of the fluidic chambers highly depends on the dielectric constant of the buffer and of the thickness of membrane separating the *cis* and *trans* fluidic reservoirs. A remarkable property of interfacial nanopores is that membranes up to micrometer in thickness can be used, without significantly affecting the nanopore bionics. Additionally, the use of a borosilicate-glass support with millimeter thickness also lowers the capacitance across the sample: the high frequency noise of the interfacial nanopores is at least one order of magnitude lower compared to conventional nanopores. Yet similar to that of the long channel SiNx nanopores, the maximum low frequency noise of interfacial nanopores is considerably lower than the one in 2D nanopores: the normalized PSD measured at 1 Hz with the current squared $\left(C_{1Hz} = \frac{S_{I,1Hz}}{I^2}\right)$ for twelve different interfacial nanopores at 100mV transmembrane potential shows a normal distribution centered at $C_{1Hz} = 1.7 \times 10^{-7}$, well comparable to (~4 times higher than) in SiNx nanopores ($4.4 \times 10^{-8}$) and almost 40 times lower than in 2D nanopores ($6.3 \times 10^{-6}$) if measured under similar conditions[22] (Figure 4b). In fact, evaluating $C_{1Hz}$ is a common approach to compare the noise among different nanopore devices [22].

At frequencies below 1 kHz, a wide variety of nanoscale devices exhibit flicker noise [32], characterized by PSDs exponentially decaying with the frequency: $S_I \propto \frac{1}{f^\alpha}$. For the majority of the nanopores studied so far[22,32,33], $\alpha = 1$, hence the low frequency noise is dubbed as $1/f$ noise. At commonly used transmembrane potentials (≤200 mV), however, the PSD in the interfacial nanopores, surprisingly exhibits a stronger dependency on frequency as $1/f^2$ (i.e. $\alpha = 2$). Considerably increasing the potential, however, invokes the $1/f$ noise characteristics in the interfacial nanopores (Figure 4c). As the origin of the $1/f$ noise in conventional nanopores is yet unclear [1,22], understanding the factors altering the noise-frequency dependency in interfacial nanopores are complex, a fortiori ( Supplementary information S7 proposes few scenarios as the origin of the observed behavior).

The low frequency noise in solid-state and biological nanopores obeys Hooge's empirical relation [22,33–35] in which the normalized PSD is inversely proportional to the number of charge carriers, $C_{1Hz} \propto N^{-1}$. The model, however, ceases to explain the low frequency noise in graphene [22] and in interfacial nanopores (Figure S7c). We collected $S_I(f = 1Hz)$ for 19 different samples (with diverse $a$ and $h$ values) at various KCl concentrations and plotted against the corresponding squared currents upon applying a constant 40 mV transmembrane potential (Figure S7a-b). Interestingly, the data corresponding to each concentration level (regardless of the geometry) follows the lines of certain slopes that can be best fitted by $\frac{S_{I,1Hz}}{I^2} \propto N^{-0.65 \pm 0.05}$. The measured dependency is weaker than the Hoog's prediction, yet stronger than what was observed for graphene nanopores ($\propto N^{-0.27}$) [22].

In summary, nanopore sensors lacking a capillary depth showed the successful detection of translocating DNA molecules. Compared to the different nanopores studied so far, interfacial nanopores combine an absolute minimal channel length with outstanding mechanical stability, minimum noise level, and reduced translocation rates. The fabrication of interfacial nanopores is scalable and does not require high-level precision. Furthermore, taking advantage of the two nanogaps as potential masks directly aligned with a nanopore, the sandwiching of 2D materials in between the slabs will allow the realization of – for example – graphene nanogap [36] electrodes in a straightforward manner. Future improvements focusing on reducing even further the nanogap widths with alternative parallelepipedic templates will provide insights into sequencing applications with tunneling currents, an application never achieved hitherto, primarily because of the challenging nanofabrication considerations.

**Acknowledgments**

The work leading to this article has gratefully received funding from the European Research Council under the European Union's Seventh Framework Programme (FP/2007-2013)/ERC Grant Agreement n. 335879 project acronym 'Biographene', the FP7 funded DECATHLON Grant agreement n. 613908 'DEvelopment of Cost efficient Advanced DNA-based methods for specific Traceability issues and High Level On-site applicatioNs', the Netherlands Organization for Scientific Research (Vidi 723.013.007), and the Swedish Research Council (621-2014-6300). The authors also thank Dr. Wangyang Fu for his valuable comments on the noise analysis section.

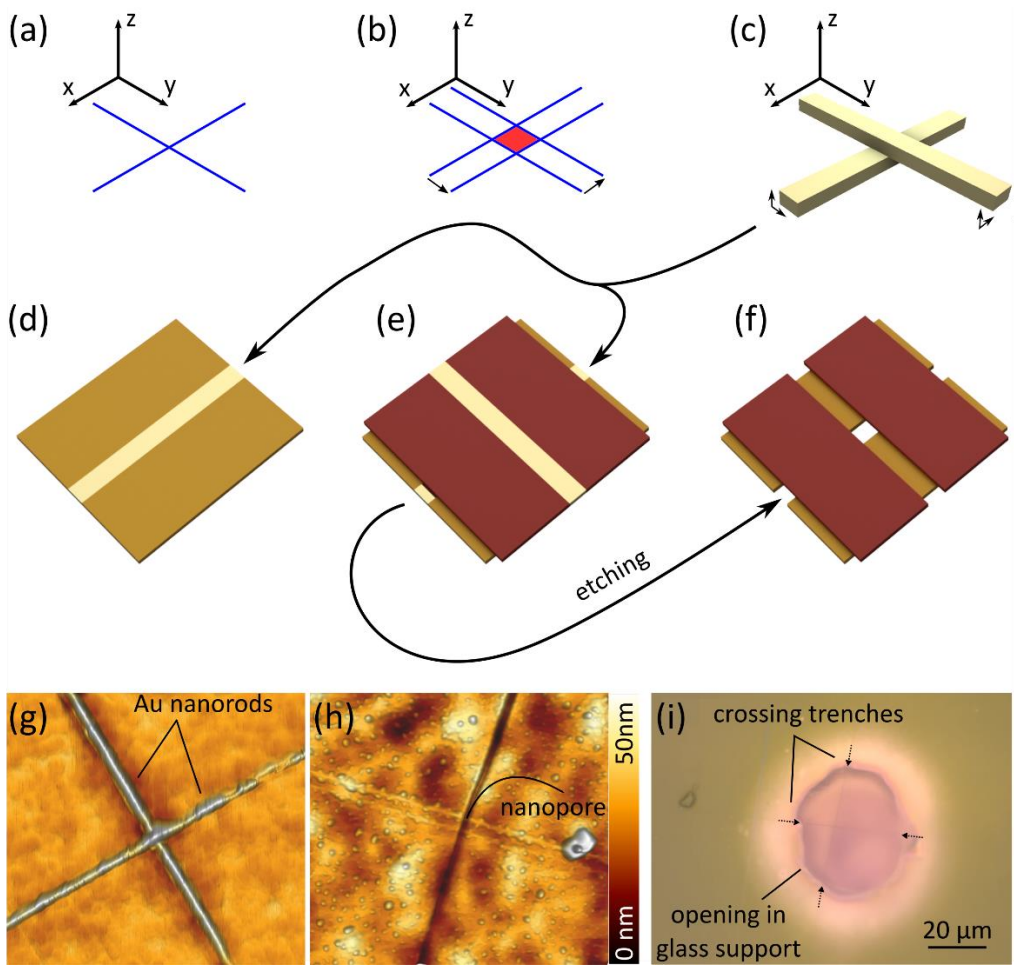

**Figure 1. Interfacial nanopores: from geometrical concepts to fabrication. a**) Zero-dimensional point at the cross-section of two crossing lines. **b**) Two-dimensional lozenge formed at the intersection of two crossing rectangles. **c**) The lozenge surface is preserved at the interface of two crossing rectangular parallelepipeds. **d**) A polymeric slab containing a parallelepipedic gold nanorod. **e**) Stack of two tangent slabs, in a twisted configuration, each containing a rectangular gold parallelepiped nanorod. **f**) Selective etching of the gold nanorods with potassium cyanide yielding an interfacial nanopore at the lozenges' interface between the slabs. **g**) Atomic force microscopy micrograph of a two slab stack showing the two tangent-crossing nanorods embedded in the polymeric matrix. **h**) Atomic force microscopy micrograph of the two slab stack after the etching of the gold using potassium cyanide. The black arrow points towards the nanopore created after the selective etching of the gold nanorods. Both the mappings in g and h are of 3μm×3μm in size. **i**) Optical micrograph of the final structure of a nanopore composed of two slabs, freely standing at the opening of a glass substrate (purple area). The dotted arrows show the lines of the two crossing parallelepipedic trenches.

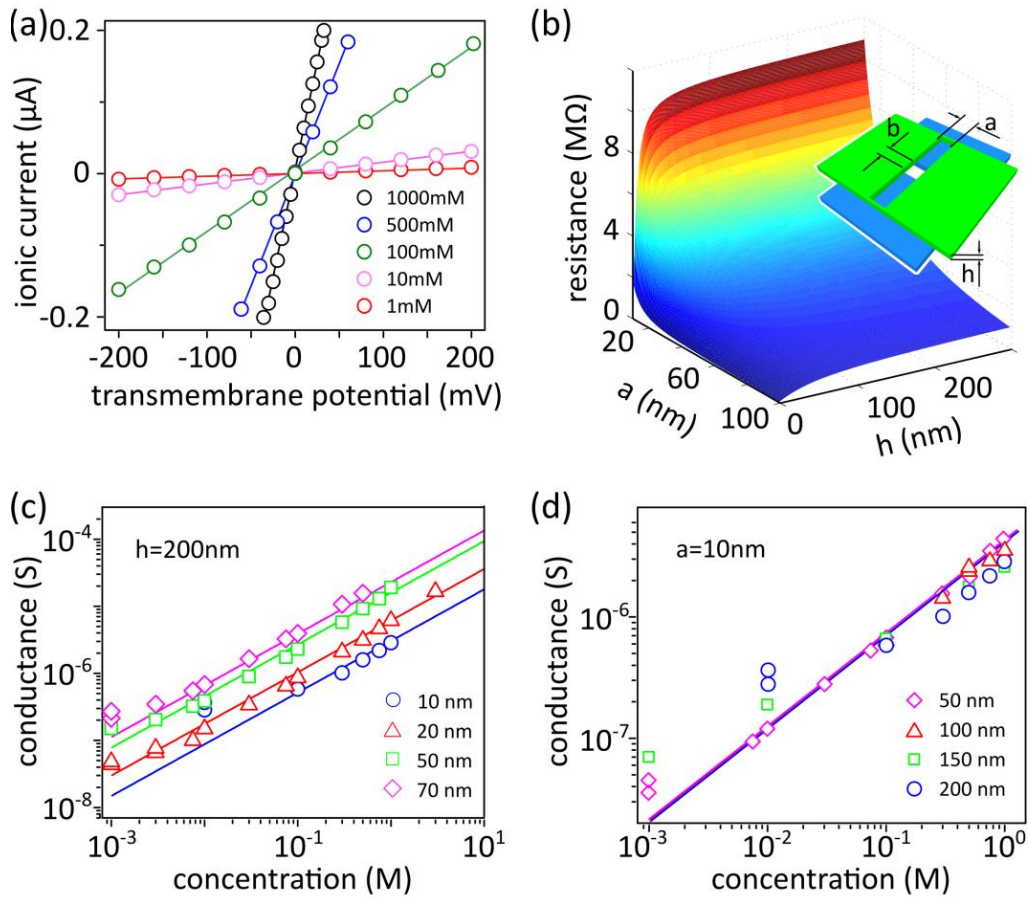

**Figure 2. Ionic transport through interfacial nanopores. a)** Ionic conduction through an interfacial nanopore ($h$=200nm, $a$=50nm) upon applying transmembrane potentials in KCl containing buffer solutions of different concentrations. **b)** Theoretically predicted resistance of interfacial nanopores as a function of the trench widths ($a=b$) and the equal thickness of the slabs ($h$). The inset depicts the three-dimensional architecture of the interfacial nanopore slab stack. The misorientation angle in between the trenches and the KCl concentration respectively were set to 90° and 1M in this mapping. **c)** Conductivity of nanopores of different trench widths ($a$) as a function of the KCl concentration; Slabs of $h$ = 200nm thickness were used to fabricate these nanopores. The continuous lines show the best fittings with equation 1. **d)** Conductivity of nanopores of different thicknesses of the slabs ($h$) as a function of the KCl concentration; All the samples were of the same trench width of $a$ = 10nm. The continuous lines show the prediction of our model for the conduction and overlap each other.

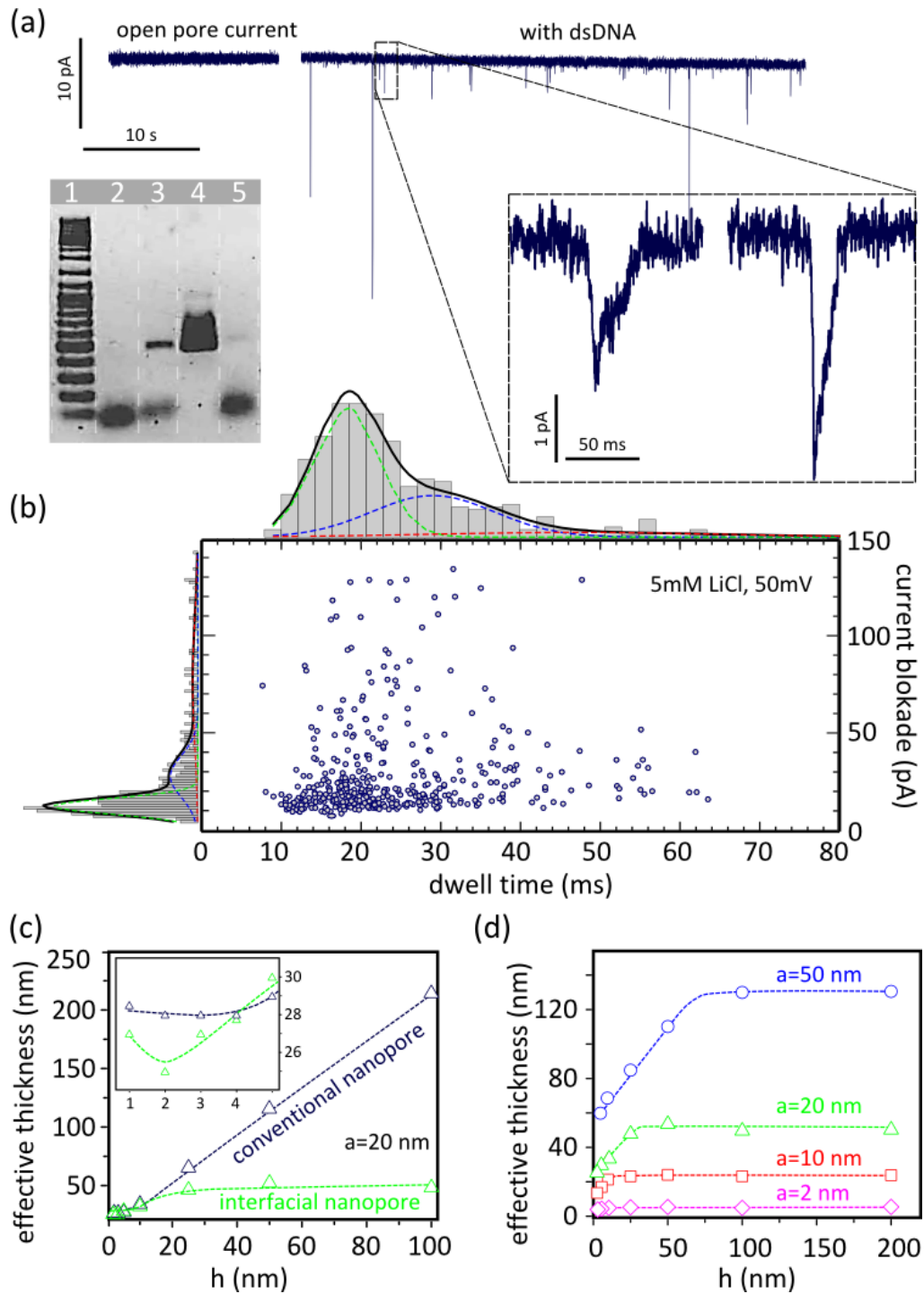

**Figure 3. Interfacial nanopore as a single molecule sensor. a)** Time-trace of the ionic current before and after the injection of λ-DNA (48.5kbp) to the cis chamber of an interfacial nanopore (*a*=50nm, *h*=50nm). The measurement is done in 5mM LiCl buffer solution and the trace is plotted after applying a low-pass filter ($f_{th} = 1kHz$). The right inset zooms on two translocation events. The left inset shows the result of the polymerase chain reaction (PCR) experiment where lanes 1, 2, 3, 4, and 5

respectively refer to the DNA marker, lambda DNA present in trans chamber before the translocation, lambda DNA present in trans chamber after translocation, 3 pg λ-DNA and water used as positive and negative controls for PCR (see supplementary information S4 for the experimental details). b) Scatter diagram of the amplitude of the current blockade versus translocation time for DNA translocation events through the same nanopore as in (**a**). The distributions of the dwell time and current blockade are separately plotted in left and top inset panels. The dashed lines represent the fits of the events to Gaussian functions. **c)** Comparison of the calculated effective thickness of interfacial and conventional nanopores at different thicknesses: The membrane thickness of the conventional nanopore is *2h* to be comparable with interfacial nanopore formed by stacking two membranes, each having the thickness *h*. Both the nanopores are of squared shape openings of $20 nm \times 20 nm$. The inset focuses on a small window at very low *h*. The vertical and horizontal axis of the inset figure have the same unit of the main panel. **d)** Evolution of the effective thickness of interfacial nanopores calculated for different slab thicknesses *h* and nanopore diameters *a*.

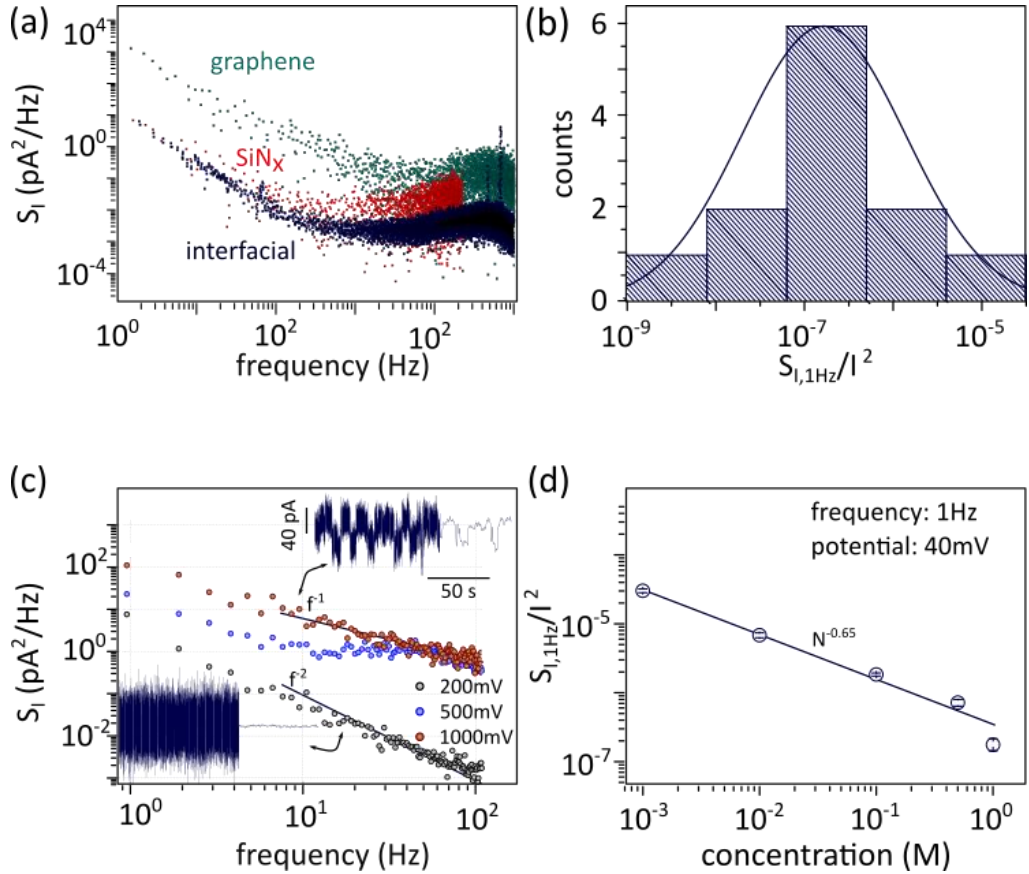

**Figure 4. Characterization of the noise in interfacial nanopores. a)** Comparison of the noise power spectral densities (PSD) of nanopores in graphene ($d = 14.2nm, R = 9.1MΩ$), in SiNx membrane ($d = 20\ nm, t = 30nm, R = 7.5MΩ$) an interfacial nanopore ($a$: $20\ nm, h$: $300nm, R = 9.9MΩ$): All the measurements performed in 1M KCl buffer solution and under 80mV transmembrane potential. **b)** Distribution of the noise power (at $f = 1Hz$) of interfacial nanopores: measurements performed with 1M KCl buffer solution and under 100mV transmembrane potential. Data from 12 different samples with diverse geometries ($100nm \leq h \leq 300nm$ and $10nm \leq a \leq 70nm$) were used. Solid line is the Gaussian fit for the distribution. **c)** Low frequency noise in an interfacial nanopore ($h = 250nm, a = 50nm$) at three different transmembrane potentials: Lines with $f^{-1}$ and $f^{-2}$ dependencies are superimposed to the data. Top and bottom insets show the corresponding signals in time-domain (right side low-pass filtered at 1kHz), respectively measured at 1V and 200mV. The same horizontal and vertical scale bars apply for both of the traces. **d)** Noise power at $f = 1Hz$ as the function of KCl concentration: The data were extracted from 19 different samples with diverse geometries ($50nm \leq h \leq 300nm$ and $10nm \leq a \leq 70nm$) under 40mV transmembrane potential (Figure S7). The dotted line shows the best fitting of the data with $C^{-0.65}$.